\documentstyle[aps,floats,prl,epsf]{revtex}
\begin{document}
\draft
\twocolumn[\hsize\textwidth\columnwidth\hsize\csname @twocolumnfalse\endcsname

\title{Classical back reaction of low-frequency cosmic 
gravitational radiation}
\author{G. Dautcourt}
\address{Max-Planck Institut f\"ur Gravitationsphysik,
Albert-Einstein-Institut,\\
Haus 5, Am M\"uhlenberg,  D~--~14476 Golm, Germany}

\maketitle

\begin{abstract}
We study in a Brill-Hartle type of approximation the back reaction
of a superposition of linear gravitational waves on its own mean
gravitational field up to second order in the wave
amplitudes. The background field is taken as a spatially flat 
Einstein-de~Sitter geometry. In order to follow inflationary scenarios,
the wavelengths are allowed to exceed the temporary Hubble distance.
As in optical coherence theory, the wave amplitudes are considered as 
random variables, which form a homogeneous and isotropic stochastic 
process, sharing the symmetries of the background metric.
A segregation of the field equations 
into equations for the wave amplitudes and equations for the background 
field is performed by averaging the field equations and interpreting 
the averaging process as a stochastic (ensemble) average.   
The spectral densities satisfy a 
system of ordinary differential 
equations. The effective stress-energy tensor for the random 
gravity waves is calculated in terms of correlation functions and 
covers subhorizon as well as superhorizon modes, where
superhorizon modes give in many cases negative contributions to energy 
density and pressure. 
We discuss solutions of the second-order equations including pure 
gravitational radiation universes.
\end{abstract}

\narrowtext
\vskip2pc]
\section{Introduction}
\label{intro}

Many classical aspects of gravitational waves are still poorly
understood. Most of them are connected with the fact that
the full theory of gravitational radiation as
following from general relativity is extremely nonlinear. Surprises
can be expected, if the nonlinear regime becomes more deeply explored.
Pure numerical methods acting on the full field equations \cite{num} 
are important but cannot be the only key to the nonlinear regime. 
They should be supplemented by suitable approximation 
methods, which allow an analytical or semi-analytical approach. 
Already in 1964 Brill and Hartle \cite{BrillHartle} proposed a 
scheme, which takes the back reaction of linear gravitational waves 
on the background metric into account. Usually, the Brill-Hartle 
method (see also \cite{BH},\cite{AB}) is considered as a 
{\it high-frequency}
approximation for gravitational radiation. For cosmological 
application the full spectrum of gravitational radiation 
including low-frequency (superhorizon) modes must be studied, 
if its generation and propagation through inflationary stages is 
considered.  We discuss in this article a different interpretation 
of the Brill-Hartle approach, which allows to treat also 
{\it low} frequencies.

The Brill-Hartle method requires an average over small scale ripples
of the geometry. How averages can be formulated consistently in 
general relativity, is a basic and still not completely solved 
problem \cite{average}.  Our assumption is to interprete 
the perturbations of the geometry as {\it random functions} as in 
optical coherence theory \cite{Wolf} or in the
Monin-Yaglom approach to statistical fluid mechanics \cite{Monin}.
The averages of the Brill-Hartle method are then taken as ensemble 
averages, not as averages over space-time regions. How stochastic 
averaging and a random geometry can be reconciled with general 
relativity in a rigorous way is also an open question. The problem 
is not touched upon here, we adopt instead a field theoretical approach 
in the sense of the Monin-Yaglom treatment.
 
Along this line we discuss the response of the expansion rate 
and of the wave amplitudes to the mean gravitational field 
produced by the waves.
We assume that the stochastic process for the wave amplitudes 
is homogeneous and isotropic. It is not necessary to suppose 
it to be Gaussian, since only two-point correlation functions are 
involved at the approximation level considered here. The background 
is kept as simple as possible, a flat 3-geometry with scale factor 
$a(\eta)$ and  conformal time $\eta$ is assumed. The treatment 
is completely classical, we do not consider back reaction effects 
from quantum gravity \cite{Tsamis}, semiclassical gravity 
\cite{squant} or an Einstein-Langevin equation \cite{Hu}. 
With methods similar to those discussed here,
the subject was treated in \cite{Abramo1} and \cite{Abramo2}. 
 
The paper is organized as follows.
In section II the segregation of the field
equations into equations for the wave amplitudes and equations for 
the background field is discussed and, following  earlier 
treatments \cite{Daut1},\cite{Daut2},
the effective stress-energy tensor for the random gravity waves
is calculated in terms of correlation functions for the wave 
amplitudes. Depending on the wave spectrum, major contributions to 
this tensor may come from waves with wavelengths exceeding the 
horizon distance.  Section III is concerned with gauge problems.
The rest of the paper treats solutions of the averaged field 
equations with various assumptions for the spectrum of gravitational 
waves and for the presence of matter fields. The aim is to 
discuss effects of back reaction in general, we do not consider the 
origin of the waves and will also not take observational constraints 
into account in this article. 
In section IV we consider the evolution of the 
gravitational wave amplitudes for different frequency regimes in 
cosmological models, using a number of simplifications. 
Pure gravitational wave models ("geons") are the subject of 
Section V. Its discussion is complicated by the necessity 
to include a non-linear back reaction term in the wave equation.  
In the final section VI possible 
improvements of the approach are shortly discussed.

\section{Averaging the field equations.}
\label{fieldeqns}

We take an Einstein-de~Sitter model as background metric
and add tensor perturbations, writing\footnote{We follow 
the conventions by Thorne, Misner and Wheeler \cite{MTW}.}
$(i,k =1,2,3)$
\begin{equation}
ds^2 = -a(\eta)^2d\eta^2 + g_{ik}dx^idx^k
\label{metric} \end{equation}
with
\begin{equation}
g_{ik} = a^2\delta_{ik} + h_{ik},~~h_{ii} = 0,~~h_{ik,k} = 0. 
\label{simp}
\end{equation}
It is appropriate to write the field equations to this metric in the
(3+1) form, assuming a perfect fluid as matter in the background 
metric:
\begin{eqnarray}
R^{(3)}+ K^2 -K_{ik}K^{ik} &=& 16\pi G\rho_m, \label{r00}\\
g^{kl}K_{ki|l} -K_{,i} &=& 0, \\
-K_{ik}'/a + R^{(3)}_{ik}+KK_{ik}-2K_{il}K_k^{~l} &=&
\nonumber \\
 4\pi G g_{ik}(\rho_m-p_m).  && \label{rik}
 \end{eqnarray}
The covariant derivatives are denoted by a prime and taken with 
respect to $g_{ik}$, and $' = \frac{d}{d\eta}$. Indices are moved 
with the three-metric $g_{ik}$. $R^{(3)}_{ik}$ and $R^{(3)}$ are 
the three-dimensional Ricci tensor and scalar for $g_{ik}$
The extrinsic curvature $K_{ik}$ is given by
\begin{equation} K_{ik} \equiv - \frac{1}{2a}g_{ik}'= -a' \delta_{ik}
-\frac{1}{2a}h_{ik}'. \end{equation}
Up to second order we have
\begin{equation} g^{ik} = \delta_{ik}/a^2 -h_{ik}/a^4 
+ h_{il}h_{kl}/a^6.
\label{up} \end{equation}

With the decompositions (\ref{simp}) and (\ref{up}) one can write 
down the field equations explicitly, including all terms up to 
the second order. Simplification results from the gauge 
restrictions for $h_{ik}$, which are taken 
into account. The Ricci tensor $R^{(3)}_{ik}$ is given by
\begin{eqnarray} & &
R^{(3)}_{ik} = -\frac{1}{2a^2}h_{ik,ll} \nonumber \\ & &
+ \frac{1}{2a^4}h_{lm}(h_{ik,lm} +
h_{lm,ik} -h_{il,km} -h_{kl,im}) \nonumber \\
& & + \frac{1}{4a^4}h_{lm,i}h_{lm,k}
+ \frac{1}{2a^4}( h_{il,m}h_{kl,m}-h_{il,m}h_{km,l}),
\label{ric3} \end{eqnarray}
and the field equations read explicitly
\begin{eqnarray}& &
6\frac{a'^2}{a^4} \nonumber \\ && + \frac{1}{a^6}(h_{kl}h_{kl,mm}
+ \frac{3}{4}h_{kl,m}h_{kl,m} -\frac{1}{2}h_{kl,m}h_{km,l})  
\nonumber \\&&
-\frac{a'}{a^7}h_{kl}h'_{kl}+3\frac{a'^2}{a^8}h_{kl}h_{kl}
-\frac{1}{4a^6}h'_{kl}h'_{kl} =  16\pi G\rho_m,  \label{f0}
\end{eqnarray}
\begin{eqnarray}
-\frac{3a'}{2a^6}h_{kl}h_{kl,i} -\frac{a'}{a^6}h_{kl}h_{ik,l}
+ \frac{1}{2a^5}h_{kl}h'_{ik,l} & &  \nonumber \\
- \frac{1}{2a^5}h_{kl}h'_{kl,i}
+ \frac{1}{4a^5}h'_{kl}h_{kl,i} -\frac{1}{2a^5}h_{kl,i}h'_{kl}
&= &0, \label{fi}
\end{eqnarray}
\begin{eqnarray}
(\frac{a''}{a}+\frac{a'^2}{a^2})\delta_{ik}+ \frac{1}{2a^2}h''_{ik}
-\frac{a'}{a^3}h'_{ik} +2\frac{a'^2}{a^4}h_{ik} & &  \nonumber \\
-2\frac{a'^2}{a^6}h_{il}h_{kl} -\frac{1}{2a^4}h'_{il}h'_{kl}
+\frac{a'}{a^5}h'_{il}h_{kl} +\frac{a'}{a^5}h'_{kl}h_{il} & &  
\nonumber \\
-\frac{a'}{2a^5}\delta_{ik}h'_{lm}h_{lm}
+ \frac{a'^2}{a^6}\delta_{ik}h_{lm}h_{lm}+ R^{(3)}_{ik}& & 
\nonumber \\
 = 4\pi G g_{ik}(\rho_m -p_m)& & . \label{fik}
\end{eqnarray}
The Brill-Hartle method usually starts with the assumption that the
space-time variation of the small ripples 
$h_{\mu\nu}$ (of order $h/\lambda$, where $\lambda$  is a typical 
radiation wavelength) is much larger than the variation of the 
background metric (of order $1/L$).
Thus terms in the Ricci tensor which are bilinear in 
$h_{\mu\nu,\rho}$ (of order $\frac{h^2}{\lambda^2}$) are comparable 
to terms involving the background metric (of order $\frac{1}{L^2}$), 
if $h \approx \frac{\lambda}{L}$. This latter condition excludes 
low-frequency waves $\lambda \ge L$, since $h$ must be sufficiently 
small to allow a second-order approximation. Terms linear in 
$h_{\mu\nu,\rho\sigma}$ are much larger and should therefore vanish 
separately, giving rise to the linear wave equation for 
$h_{\mu\nu}$. 

We do not follow this bookkeeping ( it was criticized in \cite{CFP}).
Instead, the functions $h_{ik}$ are interpreted as {\it random 
functions}. Performing 
a stochastic average of (\ref{f0},\ref{fi},\ref{fik}) removes terms 
linear in $h_{ik}$, but keeps terms bilinear in $h_{ik}$ and in the 
derivatives of $h_{ik}$. This allows a segregation of the field 
equations without restrictions for the wavelengths. The result of the 
stochastic average can 
formally be written as (Eqn.(\ref{fi}) reduces to an identity)
\begin{eqnarray}
3\frac{a'^{2}}{a^4}  &=& 8\pi G(\rho_m +\rho_g), \label{g0}\\
 \frac{a''}{a}+ \frac{a'^{2}}{a^2}
 &=& 4\pi G a^2(\rho_m -p_m +\rho_g - p_g), \label{gik}
\end{eqnarray}
where $\rho_g$ and $p_g$ 
are the averages over nonlinear terms.
It is convenient to interprete $\rho_g,p_g$ as the effective density 
and pressure of the gravitational radiation field.
Subtracting the averaged field equation (\ref{gik}) (multiplied with
$\delta_{ik}$) from (\ref{fik}) and neglecting higher-order terms of 
the type $\langle A\rangle -A$, as is usually done in a 
self-consistent field approximation, one obtains a {\it modified}    
wave equation for the amplitudes $h_{ik}$
\begin{equation}
h_{ik}^{''} - \Delta h_{ik} - 2\frac{a'}{a}h^{'}_{ik}
+ 2h_{ik}(\frac{a'^2}{a^2} -\frac{a^{''}}{a}+b)
=0 \label{we}
\end{equation}
with 
\begin{equation}
b = 4\pi\zeta Ga^2(\rho_g-p_g).
\end{equation}
This equation differs from the usual form 
by a time-dependent term $b$
representing a back reaction of the energy and pressure of the waves 
on their propagation (a factor of $\zeta$ was introduced in 
front of $b$ in order to switch off the back reaction term  
for comparison purposes). Since $\rho_g$ and $p_g$ depend on 
solutions of the wave equation, some nonlinearity is thus
introduced. $b$ results from second-order bilinear terms and
is therefore neglected in linear treatments of cosmic wave 
propagation, but should be kept in the spirit of our approach.  
In situations where the waves do not appreciably influence 
the scale factor evolution, the back reaction term can be neglected. 
Also its influence is usually small in the high-frequency
approximation, when the wavelengths are small compared to
the Hubble distance.
Note also that the back reaction term is exactly zero, if the wave 
background has the Zeldovich equation of state $p_p =\rho_g$. 
On the other hand, if low-frequency radiation contributes 
appreciably to the average density and pressure and hence to the
scale factor evolution, the 
modified wave equation must be considered in general.            
(A previous paper (\cite{Daut3}) on the same subject was based on the
linear wave equation with $b=0$. We shall continue its 
use in Sec. IV in order to compare later the results with those 
of the general case $b \neq 0$).

Some authors use different definitions of the wave amplitudes
by applying factors of $a$ on them. We have defined the spatial
components $h_{ik}$ as perturbations to the three-metric $a^2\delta_{ik}$.
Equivalent to (\ref{we}) are
\begin{equation}
(h_{ik}/a^2)^{''} - \Delta (h_{ik}/a^2)
+ 2\frac{a'}{a}(h_{ik}/a^2)^{'} +2 b h_{ik}/a^2 = 0
\label{we1}
\end{equation}
and
\begin{equation}
(h_{ik}/a)^{''} - \Delta (h_{ik}/a)
+(2b -\frac{a''}{a})(h_{ik}/a)= 0,
\label{we2}
\end{equation}
which are sometimes easier to use. It was frequently noted that
gravitational wave perturbations in a Friedman-Robertson-Walker (FRW) universe
may be described as a pair of massless minimally coupled scalar fields
in  the same background space-time (see, e.g., Ford and Parker in 
(\cite{squant}).
If back reaction is important, this correspondence is lost in general,
but for a de~Sitter scale factor
$a \sim 1/\eta $ the modified wave equation has the form
$(\Box +\xi R +m^2)(\Phi/a)=0$ with 
$\xi= \frac{1}{2}-\frac{1}{6}m^2a^2\eta^2$, which characterizes
non-minimally coupled and possibly massive scalar 
particles.

To calculate the components of the gravitational stress-energy 
tensor, we assume that the random process represented by $h_{ik}$ 
is homogeneous and isotropic. The correlation functions which 
enter (\ref{g0},\ref{gik}) are 
related to certain spectral densities. To see this, we represent  
$h_{ik}$ as stochastic Fourier integral (see \cite{Monin} 
for a detailed treatment of the spectral representation of random 
processes):
\begin{equation}
h_{ik}({\bf x},\eta) = \int \gamma_{ik}({\bf k},\eta)
e^{i{\bf k}{\bf x}}d{\bf k} + conj.compl.
\end{equation}
From (\ref{we}), the Fourier amplitudes satisfy the ordinary 
differential
equation
\begin{equation}
\gamma_{ik}^{''}
-2\frac{a'}{a} \gamma^{'}_{ik} + \gamma_{ik} (2\frac{a'^2}{a^2}
-2\frac{a''}{a} + k^2+ 2b) =0. \label{four1}
\end{equation}
Amplitudes of the correlation functions may now be written
as frequency integrals over spectral densities:
\begin{equation}
\langle h_{ik}({\bf x},\eta)h_{lm}({\bf x},\eta)\rangle
=  \int\langle  (\gamma_{iklm}+\gamma^{*}_{iklm})\rangle d{\bf k},
\label{amp1}
\end{equation}
where\footnote{We refer to equal space - equal time correlators          
of the form (\ref{amp1}) as "correlation functions".}
\cite{Monin}
\begin{eqnarray}
\langle \gamma_{ik}({\bf k},\eta)\gamma^{*}_{lm}(\tilde{\bf k},
\eta)\rangle  &
 = & \delta({\bf k} - \tilde{\bf k})\gamma_{iklm}, \label{spec1}\\
\langle \gamma_{ik}({\bf k},\eta)\gamma_{lm}(\tilde{\bf k},
\eta)\rangle &=& 0.
\end{eqnarray}
It is easy to see that the spectral densities $\gamma_{iklm}$ satisfy 
the symmetry relations
\begin{eqnarray}
 \gamma^{*}_{iklm} & = &  \gamma_{lmik}, ~  \gamma_{iklm}
=  \gamma_{kilm}, \label{cons1}\\
 \gamma_{ikll} & = & 0, ~~ \gamma_{iklm}k^m   =  0. \label{cons2}
\end{eqnarray}
The general solution of the algebraic constraints (\ref{cons1},
\ref{cons2}) for the spectral densities contains four complex 
functions of $k$ and $\eta$, which describe polarized background 
radiation in general. The stochastic background of gravitational 
waves expected from pre-galactic stages of the Universe could be 
polarized due to strong anisotropies expected at Planck time.  
Here we confine the discussion to unpolarized radiation, represented 
by a single (real) spectral function
$\alpha(k,\eta)$. 
Then $\gamma_{iklm}$
can be written as:
\begin{eqnarray}
\gamma_{iklm} & = & \alpha(k,\eta) \bar\delta_{iklm}, \\
\bar\delta_{iklm} & = & \bar\delta_{il}\bar\delta_{km}
+\bar\delta_{im}\bar\delta_{kl}-\bar\delta_{ik}\bar\delta_{lm},
\end{eqnarray}
where
\begin{equation}
\bar\delta_{ik} =\delta_{ik}-k_ik_k/k^2,~~ k^2=k^lk^l.
\end{equation}
The transversal $\delta$-symbol satisfies
\begin{equation}
\bar\delta_{ii}=2, ~~ \bar\delta_{ik}k^k =0,~~
 \bar\delta_{il} \bar\delta_{kl} = \bar\delta_{ik}.
\end{equation}
To calculate the averaged stress-energy tensor, we need also the 
spectral densities
\begin{eqnarray}
\langle \gamma'_{ik}\gamma'^{*}_{lm}\rangle
&=& \beta \delta({\bf k} - \tilde{\bf k}) \bar \delta_{iklm}, 
\label{spec2}\\
\langle \gamma_{ik}\gamma'^{*}_{lm}\rangle
&=& \gamma \delta({\bf k} - \tilde{\bf k}) \bar \delta_{iklm}, 
\label{spec3}
\end{eqnarray}
where $\beta$ is real and $\gamma$ is complex in 
general (the correlation functions $\alpha$, $\beta$ and $\gamma$ 
were introduced in \cite{Daut2}). Similar to
(\ref{amp1}), the correlation functions which are needed to calculate 
the effective wave stress-energy tensor can be written as
\begin{eqnarray} \langle h_{ik,r}h_{lm,s}\rangle  &=&
2 \int \alpha\bar{\delta}_{iklm}k_{r}k_{s}d{\bf k}, \nonumber \\
\langle h_{ik}h_{lm,rs}\rangle  &=& -\langle h_{ik,r}h_{lm,s}\rangle , 
\nonumber \\
\langle h'_{ik}h'_{lm}\rangle  &=& 2\int \beta \bar{\delta}_{iklm}
d{\bf k},\nonumber \\
\langle h_{ik}h'_{lm}\rangle  &=& 2\int \Re(\gamma)\bar{\delta}_{iklm}
d{\bf k},\nonumber\\
 \langle h_{ik}h_{lm,r}\rangle  &=&
2 \int \alpha k_r \bar{\delta}_{iklm}d{\bf k},\nonumber \\
 \langle h_{ik}h'_{lm,r}\rangle  &=&
2 \int \Im(\gamma)\bar{\delta}_{iklm}k_{r}d{\bf k}\nonumber\\
\langle h'_{ik}h_{lm,r}\rangle  &=& -\langle h_{ik}h'_{lm,r}\rangle.
\nonumber
\end{eqnarray}
Since $\alpha,\beta,\gamma$ are functions of $k$ (and $\eta)$ only, 
the angular
integrations can be performed easily.  This gives with  $p_{iklm}
\equiv \delta_{il}\delta_{km}+\delta_{im}\delta_{kl}$:
\begin{eqnarray}
\langle h_{ik}h_{lm}\rangle
&=& \frac{16\pi}{15}(3p_{iklm}-2\delta_{ik}\delta_{lm})
\int \alpha k^2 dk,\nonumber\\
\langle h_{ik,r}h_{lm,s}\rangle   &=&
\frac{16\pi}{105}(-10\delta_{ik}\delta_{lm}\delta_{rs}+11 
p_{iklm}\delta_{rs} \nonumber\\ &&
-3p_{ilrs}\delta_{km}-3p_{lkrs}\delta_{im} \nonumber \\ &&
-3p_{kmrs}\delta_{il}-3p_{imrs}\delta_{kl} \nonumber \\ &&
+4p_{ikrs}\delta_{lm}+4p_{lmrs}\delta_{ik})
\int \alpha k^4 dk, \nonumber\\
\langle h_{ik}h_{lm,rs}\rangle  &=& -\langle h_{ik,r}h_{lm,s}\rangle 
\nonumber\\
\langle h'_{ik}h'_{lm}\rangle
&=& \frac{16\pi}{15}(3p_{iklm}-2\delta_{ik}\delta_{lm})
\int \beta k^2 dk, \nonumber \\
\langle h_{ik}h'_{lm}\rangle  &=&
\frac{16\pi}{15}(3p_{iklm}-2\delta_{ik}\delta_{lm})
\int \Re(\gamma)k^2 dk, \nonumber\\
\langle h_{ik}h_{lm,r}\rangle    &=& 0,\nonumber \\
\langle h'_{ik}h_{lm,r}\rangle   &=& 0,\nonumber \\
\langle h_{ik}h'_{lm,r}\rangle   &=& 0.\nonumber
\end{eqnarray}
Since only the real part of $\gamma$ enters the averaged
expressions, we denote subsequently this real part by $\gamma$. 
Some unusual results can be expected when one deals with stochastic 
averages over nonlinear quantities. For the averaged 
three-dimensional Ricci tensor we obtain, using the just derived 
relations,
\begin{equation} \langle R_{ik}^{(3)}\rangle
= \frac{8\pi}{3a^4}\delta_{ik} \int \alpha k^4dk.
\end{equation}
Thus, whereas the three-dimensional background metric is flat, a 
random superposition of gravitational radiation to this background 
produces an averaged 3-Ricci tensor with positive curvature, if 
curvatures are defined as eigenvalues of the Ricci tensor. On the 
other hand, the stochastic average of the 3-curvature scalar 
$R^{(3)} =g^{ik}R^{(3)}_{ik}$ is negative:
\begin{equation} \langle R^{(3)}\rangle  = -\frac{8\pi}{a^6} 
\int \alpha k^4dk.
\end{equation}
The bilinear terms which enter $R^{(3)}$ and produce a nonzero 
stochastic average are different from those entering $R^{(3)}_{ik}$, 
thus the result $\langle R_{ik}\rangle \langle g^{ik}\rangle
 \neq  \langle R_{ik}g^{ik}\rangle$ does not come as a surprise.  But
it makes it hard to interprete stochastic averages geometrically.

Energy density and pressure of the gravitational waves are given by 
similar integrals
\begin{eqnarray}
\rho_{g} &=& \frac{1}{2Ga^6} \int dk k^2 (k^2\alpha +\beta
+ 4\gamma \frac{a'}{a}  -12\alpha \frac{a'^2}{a^2}), \label{rh}\\
p_{g} &=& \frac{1}{6Ga^6} \int dk k^2 (7k^2\alpha -5\beta
+ 20\gamma \frac{a'}{a} -20\alpha \frac{a'^2}{a^2}). \label{pr}
\end{eqnarray}
The deviation from the HF equation of state follows as
\begin{equation}
\rho_{g} -3p_{g} = \frac{1}{Ga^6} \int dk k^2 (-3k^2\alpha +3\beta
- 8\gamma \frac{a'}{a}  + 4\alpha \frac{a'^2}{a^2}), \label{dev}\\
\end{equation}
and the back reaction function $b$ is given by
\begin{equation}
b = \frac{8\pi\zeta}{3a^4} \int dk k^2 (-k^2\alpha +2\beta
- 2\gamma \frac{a'}{a}  -4\alpha \frac{a'^2}{a^2}). \label{br}\\
\end{equation}

The time evolution of the spectral functions $\alpha$, $\beta$ and 
$\gamma$ is obtained from (\ref{four1}).
Differentiating (\ref{spec1}),(\ref{spec2}),(\ref{spec3})
with respect to $\eta$ and using (\ref{four1}) leads to a system of 
coupled differential equations:
\begin{eqnarray}\alpha'& = & 2\gamma,  \label{eq1}\\
\beta'& =&  4\frac{a'}{a}\beta +2\gamma (2\frac{a''}{a}
- 2\frac{a'^2}{a^2} -2b -k^2), \label{eq2} \\
\gamma'& = & 2\frac{a'}{a}\gamma + \beta
+ \alpha(2\frac{a''}{a} -2\frac{a'^2}{a^2} -2b -k^2). \label{eq3}
\end{eqnarray}
With given initial values the time evolution of all spectral
densities and hence correlation functions follows from this system, 
if the scale factor is known. The system for $\alpha, \beta, 
\gamma$ is actually nonlinear due to the presence of the $b$ term.
It can be transformed into a single nonlinear differential
equation for $\alpha$ alone. Substituting $\alpha'/2$ for
$\gamma$ in the last two equations and introducing a new function
$\epsilon$ instead of $\beta$ by means of
\begin{equation} \epsilon = \beta -\frac{\alpha'^2}{4\alpha},
\end{equation}
one obtains
\begin{equation} \epsilon' + (\frac{\alpha'}{\alpha}
-4\frac{a'}{a})\epsilon = 0. \end{equation}
This is integrated to
\begin{equation} \epsilon =\frac{\epsilon_0(k) a^4}{\alpha}.
\end{equation}
\noindent
Then the  basic spectral function $\alpha$ satisfies the nonlinear 
differential equation
\begin{eqnarray} && 2\alpha \alpha'' - \alpha'^2
- 4\epsilon_0a^4 -4\frac{a'}{a}\alpha\alpha'  
- 4\alpha^2(2\frac{a''}{a}
-2\frac{a'^2}{a^2} -2b -k^2) \nonumber \\
&& = 0. \label{al}
\end{eqnarray}
If we define wave amplitudes with a different power of $a$, for 
instance as in (\ref{we2}), this equation can be simplified.
With $\alpha= fa^2$,
\begin{equation} 2ff'' - f'^2 +4f^2(k^2+2b -\frac{a''}{a})-4\epsilon_0
= 0. \label{al1}
\end{equation}
\noindent
This equation is also nonlinear in $f$, but its nonlinearity arises 
because we deal with the spectral density of an expression which is 
{\it quadratic} in the random wave amplitudes $h_{ik}$. The 
solutions of (\ref{al1}) are related to those of a 
differential equation, which is linear in the wave amplitudes. 
Let $h_1,h_2$ be a pair of real solutions to
\begin{equation}
h'' + h(k^2 +2b -a''/a) = 0 \label{al2}
\end{equation}
such that
\begin{equation}
h_1 h_2' -  h_2 h_1' = \sqrt{\epsilon_0},
\end{equation}
where $\epsilon_0$ is a time-independent function of $k$ only. 
Then $f = h_1^2 + h_2^2$ is a solution of (\ref{al1}).
Not surprisingly, the linear differential equation for $h$ is the 
Fourier transform of the wave equation (\ref{we2}) for a single 
realization of the random process.
We also note that the solutions of (\ref{al1}) can be reduced
to solutions $f_{hom}(k,\eta)$ of the homogeneous part of (\ref{al1})
($\epsilon_0= 0$): Any solution of the full equation (\ref{al1})
can be written in terms of a suitable function $f_{hom}$ as
\begin{equation}
f = f_{hom} + \epsilon_0(k)f_{hom}\left(\int 
\frac{d\eta }{f_{home}}\right)^2.\label{hom1}
\end{equation} 

It is convenient to write the
expressions for $\rho$ and $p$ and for other stochastic averages
in terms of four frequency-independent,
but in general time-dependent, integrals over the spectral density
$f(k,\eta)$, denoted as "moments" subsequently:
\begin{eqnarray}
f_0 &=& \int dk~k^2 \frac{\epsilon_0(k)}{f(k,\eta)}, 
~f_1 = \int dk~k^2 \frac{f'^2(k,\eta)}{f(k,\eta)}, \nonumber \\
~f_2 & =& \int dk~k^2 f(k,\eta),
~f_4 = \int dk~k^4 f(k,\eta).\label{mom}
\end{eqnarray}
In the density and pressure equation, $f_0$ and $f_1$ appear only in 
the combination
\begin{equation}
 g_4 = f_0+f_1/4. \label{momg}
\end{equation}

We also note the useful relations
\begin{eqnarray}
 g'_4  &=& -f_4' + (\frac{a''}{a} -2b)f_2', \label{s1}\\
 f_2''&=& 2g_4 +2(\frac{a''}{a}- 2b)f_2  -2f_4, \label{s2}
\end{eqnarray}
which follow from differentiating $f_1$ once and $f_2$ twice
and using the differential equation for $f$. 

For a general scale factor, energy density and pressure
may be rewritten as
\begin{eqnarray}
\rho_{g} &=& \frac{1}{2Ga^4} (f_4 +g_4 +3\frac{a'}{a}f'_2
-7\frac{a'^2}{a^2}f_2), \label{rh1}\\
3 p_{g} &=& \frac{1}{2Ga^4} (7f_4 -5g_4 +5\frac{a'}{a}f'_2
-5\frac{a'^2}{a^2}f_2). \label{pr1}
\end{eqnarray}
The back reaction function $b$ is given by 
\begin{equation}
 b = \frac{8\pi\zeta}{3a^2}(-f_4 +2g_4 +\frac{a'}{a}f'_2
- 4\frac{a'^2}{a^2}f_2), 
\label{br2}
\end{equation}
thus the relations (\ref{s1}),(\ref{s2}) for the moments become nonlinear
in general.
 
For completeness, we give the ensemble averages of some other 
geometrical quantities. The mean value of the four-dimensional Ricci 
scalar can be written as 
\begin{equation}
\langle R \rangle = \frac{6a''}{a^3} + \frac{8\pi}{a^4} (3f_4 -3g_4 
+\frac{a'^2}{a^2}f_2 + \frac{a'}{a}f'_2), \label{rs4}\\
\end{equation} 
where the first term is the background contribution. The second term 
may also be written as $8\pi G(3p_g-\rho_g)$. Since $R=0$ in vacuum, 
also the stochastic average of $R$ must vanish. Then (\ref{rs4})
is equivalent to a combination of (\ref{g0}) and (\ref{gik}) (with 
$\rho_m=p_m=0$), as one can check easily.
There are only two independent and in general nonzero components
of the averaged Riemann tensor:
\begin{eqnarray}
\langle R_{0101} \rangle &=& a'^2-aa'' +
\frac{8\pi}{3} (g_4  +\frac{a'^2}{a^2}f_2 - \frac{a'}{a}f'_2),
\label{riem1}\\
\langle R_{1212} \rangle &=& a'^2 +
\frac{4\pi}{3}(3f_4-g_4 -\frac{a'^2}{a^2}f_2 - \frac{a'}{a}f'_2).
\label{riem2}
\end{eqnarray} 
The one independent component of the averaged Weyl tensor
\begin{equation}
\langle C_{0101} \rangle = \langle C_{1212} \rangle =
\frac{16\pi}{3} (f_4  -\frac{a'^2}{a^2}f_2 + \frac{a'}{2a}f'_2).
\label{weyl}
\end{equation} 
contains, of course, no background term. All curvature components 
result from averaging nonlinear
terms, they have no direct connection to the first order perturbations
of the curvature tensor, whose average is zero. 
It is immediately seen that  
$\langle g^{\alpha\beta}\rangle \langle C_{\alpha\mu\beta\nu} \rangle$
differs from zero for some components, while, of course, the original
expression $g^{\alpha\beta} C_{\alpha\mu\beta\nu}$ and hence its 
average vanishes. Averaging the four local curvature invariants 
connected with the Weyl tensor is more complicated: Apart from the 
correlation functions listed above further expressions such as 
$\langle h_{ij,kl}h_{mn,rs} \rangle$ must be calculated, which
introduces higher-order moments such as $f_6= \int dk~k^6f(k,\eta)$.

We summarize the main results of this section. 
In order to describe the back reaction of gravitational radiation 
propagating in a flat Friedman universe upon the scale factor,
we have replaced the Einstein field equations by a system of 
equations for averaged quantities. This system consists (i) of the 
Friedman equations (\ref{g0}),(\ref{gik}) for a cosmic fluid with 
density $\rho_m$ and pressure $p_m$, together with the corresponding 
quantities $\rho_g$ and $p_g$ for gravitational radiation given by 
(\ref{rh1}),(\ref{pr1}), and (ii) of the differential equation 
(\ref{al1}) for the spectral density $f(k,\eta)$. The latter equation
may be replaced partially by differential relations for the 
moments of $f$. The aim is to solve 
these equations for the scale factor and for $f$ simultaneously.
One has to keep in mind that the results are only reliable 
if terms higher than second order in the original field equations 
can safely be neglected.

\section{Gauge transformations}
We have chosen to work in a fixed gauge. As recently noted
by Unruh \cite{Unruh},
there is no basic difference between a gauge invariant approach, where the
equations are written in terms of "gauge-independent" variables 
(see, e.g., \cite{Abramo2}), and
an approach where the gauges are fixed in some way.  
It is only important to fix the gauges completely,
leaving no room for residual gauge freedom, which could be mistaken 
as physical degree of freedom. We therefore ask for the coordinate 
transformations 
$\overline{x}^{\mu} =x^{\mu}+ \xi^{\mu}$ preserving the eight 
constraints $h_{00}=0,~h_{0i}=0,~h_{ii}=0,~h_{ik,k}=0$.
To linear order in $\xi^{\mu}$ the perturbations transform as
\begin{equation}
\overline{h}_{\mu\nu}(x^{\mu}) = h_{\mu\nu}(x^{\mu}) 
- g^{(0)}_{\mu\nu ,\rho}\xi^{\rho} - g^{(0)}_{\mu\rho} 
\xi^{\rho}_{,\nu} - g^{(0)}_{\nu\rho} \xi^{\rho}_{,\mu}. 
\end{equation}
A simple calculation shows that for a general scale factor the
transformations we are looking for must satisfy
\begin{equation}
\xi^0=0,~\xi^i = \xi^i(x^k),~\xi^i_{,i}= 0,~\xi^i_{,kk}=0.
\end{equation}
The functions $\xi^{\mu}$ therefore do not necessarily form a 
Killing field, and the first-order perturbations are gauge-dependent 
in general. The spatial components transform as 
\begin{equation}
\overline{h}_{ik}(x^i) =  h_{ik}(x^i) - a^2(\xi^i_{,k}+ \xi^k_{,i}).
\label{trans1}
\end{equation}
It is easy to see that the first-order wave equation (\ref{we}) is 
invariant with respect to (\ref{trans1}).
The transformation of the correlation functions 
is derived 
assuming that the vector $\xi^i$ is also a random process with zero
mean and independent of the wave process. We outline the procedure. 
The random process $\xi^i$ must be homogeneous and isotropic as 
$h_{ik}$, so its correlation function can be written
\begin{equation}
\langle \xi^i\xi^k \rangle = \int d{\bf k}(F_1(k)\frac{k^ik^k}{k^2}
+ F_2(k)\delta_{ik}).
\end{equation}
$\xi^i_{,i}=0$ is translated into $F_1(k)= -F_2(k)$ in the spectral
region. It is more difficult to write the condition $\xi^i_{,kk}=0$ 
directly as condition for the spectral density $F_1$, so we first 
proceed without this constraint. 
With (\ref{trans1}) one forms 
$\langle \overline{h}_{ik}\overline{h}_{lm} \rangle$, this gives
\begin{equation}
\overline{f}_2 = f_2 + \frac{a^2}{2}\int dk k^4 F_2(k).\label{tran1}
\end{equation}
Similarly, the reduction of 
$\langle \overline{h}'_{ik}\overline{h}'_{lm} \rangle$ yields       
\begin{equation}
\overline{g}_4 = g_4 + \frac{a'^2}{2}\int dk k^4 F_2(k).\label{tran2}
\end{equation}
Writing out $\langle \overline{h}_{ik,r}\overline{h}_{lm,s} \rangle$ 
shows that the average of 
the $\xi^i$-depending terms does not have the symmetries of the
other terms. This reflects the fact that the condition 
$\xi^i_{,kk}=0$ was not taken into account. We conclude that the 
spectral density satisfies the integral relation 
$\int dk k^6F_2(k)=0$, and that $f_4$ transforms as 
\begin{equation}
\overline{f}_4 = f_4. \label{tran3}
\end{equation}
It is easily seen that the density $\rho_g$ and pressure $p_g$ 
as defined by (\ref{rh}),(\ref{pr}) are invariant
against the transformations (\ref{tran1})-(\ref{tran3}),
and this holds also for the averaged components of the 
Weyl tensor and for the Ricci scalar.
 
As is well known, the effective stress-energy tensor is {\it not} gauge 
invariant with respect to {\it general} coordinate transformations, 
i.e., transformations which violate the constraints. However, as 
shown by Abramo, Brandenberger and Mukhanov (\cite{Abramo1}, 
see also \cite{Bruni},\cite{AB}), the gauges change at the same 
time the background geometry to second order, and these changes 
just compensate the change of the energy-stress tensor. 

\section{Solutions including matter fields}     
We now try to find solutions of the equations derived in section II,
assuming that apart from gravitational radiation other forms of 
matter are present, which dominate dynamically during most stages 
of the cosmic evolution, if not always. Dominance of matter means 
that the $b-$term in the wave equation can be neglected.
\subsection{The high-frequency regime}
The high-frequency (HF) regime is defined by the 
assumption that all wavelengths
are small compared with the temporary Hubble distance.
Then the time derivatives which occur in (\ref{eq1}-\ref{eq3})
are small: Let $A$ be one of the quantities $\alpha,\beta,\gamma$, 
then $A'$ is of the order $A/T$, where $T$ is a Hubble time,
$T \approx a/a'$, and this is much smaller then a multiplication with 
the wave frequency $k$: $A/T \ll Ak$. The terms $\beta$ and 
$k^2\alpha$  in (\ref{eq3}) are therefore much larger than other 
terms and must cancel in a HF approximation:
\begin{equation} \beta = k^2\alpha. \end{equation}
Using this relation and neglecting time derivatives in (\ref{rh},
\ref{pr}), one obtains
\begin{equation} \rho_g = 3p_g = \frac{1}{Ga^6}\int dk k^4\alpha,
\label{highf} \end{equation}
equivalent to the equation of state for a gas of noninteracting 
massless particles.  A similar cancellation of terms must occur in 
(\ref{al}). This requires $\epsilon_0 > 0$,  gives
 \begin{equation} \alpha = \sqrt{\epsilon_0}a^2/k \end{equation}
and identifies the function $\epsilon_0(k)$ as closely related to 
the time-independent spectral function in the HF regime.
The energy density in this regime, 
\begin{equation} \rho_g = \frac{1}{Ga^4}\int dk k^3 \sqrt{\epsilon_0},
\label{hel}
\end{equation}
shows the typical $1/a^4$ dependence on the scale factor for
background radiation. Apart from an overall redshift factor,
the gravitational wave spectrum does not change in time. 

The expression (\ref{highf}) for the effective gravitational energy 
density and pressure of {\it high-frequency} gravitational radiation 
holds also in the case that this radiation contributes appreciably 
to the background geometry, beside other forms of matter. The result
is only a changed time dependence of the scale factor $a$. 

\subsection{Stress tensor evolution in a fixed background}
Considering now the full spectrum including both low-frequency (LF) 
and HF modes, one must return to the solution of the complete equation 
(\ref{al1}). The change of the spectrum during expansion depends on 
the wavelength. The solutions $f$ of (\ref{al1}) for power law scale 
factors can be represented by Hankel functions of second kind, if
we neglect the $b-$term assuming that gravitational waves do not 
contribute appreciably to the scale factor. We 
prefer explicit expressions, since this allows to use Fourier methods. 
For the matter universe ($a \sim \eta^2$), the radiation universe 
($a \sim \eta$) and the de~Sitter universe ($a \sim 1/\eta$) as
background geometry the general solution can then be written
\begin{equation}
\alpha/a^2 = f = 2npp^* +(l+im)p^2 +(l-im)p^{*2}. \label{fall}
\end{equation}
$l,m,n$ are three functions of $k$ and connected with $\epsilon_0$
by
\begin{equation}
\epsilon_0 = 4k^2(n^2-l^2-m^2). \label{ep}
\end{equation}
$p(x)$ is a complex function of $x= k\eta$, given by
\begin{equation}
p(x) = (1+is/x)e^{ix}  \label{frad}
\end{equation}
with $s=1$ in the matter and de~Sitter universe and $s=0$ in the
radiation universe.
(In spite of the different scale factors, the spectral density $f$ 
in the de~Sitter cosmos has the same form as in the matter universe, 
since $a''/a$ is the same. This coincidence holds only for the 
spectral density $f$ and is destroyed, if one returns to $\alpha$). 
For large $x$, $f$ reaches the same asymptotic form in all models.
This asymptotic form coincides with the HF expression 
$f=\sqrt{\epsilon_0}/k$ considered previously, if the ratios 
$\frac{l}{n}$ and $\frac{m}{n}$ tend to zero for large $k$.

How do the different wavelengths contribute to the effective 
stress-energy tensor?  It is usually assumed that modes with 
wavelengths larger than the temporary Hubble radius can be discarded, 
since they are unobservable and not true waves \cite{Zel}, have no 
appreciable influence on the local wave energy density \cite{Allen1},
\cite{Tavares}, \cite{Garcia}, or appear locally as gauge 
transformation \cite{Sahni},\cite{Unruh}. The relations (\ref{rh1}) 
and (\ref{pr1}) allow a definite answer. For a radiation universe 
$a \sim \eta$ one obtains:
\begin{eqnarray}
a^4 G \rho_{g} &=& 2n_4 -\frac{7}{\eta^2}n_2
-\frac{7}{2\eta^2}\psi_2(\eta)+\frac{3}{2\eta}\psi'_2(\eta),
\label{rhs}\\
3a^4 G p_{g} &=&  2n_4 -\frac{5}{\eta^2}n_2-\frac{5}{2\eta^2}
\psi_2(\eta) +\frac{5}{2\eta}\psi'_2(\eta) \nonumber \\
&& - \frac{3}{2} \psi''_2(\eta) \label{prs}
\end{eqnarray}
with the time function  
\begin{equation}
\psi_2(\eta) = 2\int^\infty_0 k^2(l(k)cos(2k\eta)-m(k)sin(2k\eta))dk,
\end{equation}
we also use the notation 
$ n_l = \int^\infty_0 n(k)k^l dk$
for the time-independent moments of $n(k)$.
The spectral functions $l,m,n$ are constrained 
by the condition that the frequency integrals (including the moments
$n_2$ and $n_4$) converge. To justify the assumption of a 
pre-determined scale factor, the amplitudes must be small enough 
that their contribution to the background geometry is negligible. 
If the function $\psi_2(\eta)$ with its first time derivative is
bounded for large $\eta$ and $\psi''_2$ tends to zero for large 
$\eta$, the relation $p_g = \rho_g/3$ between
pressure and density follows asymptotically for large $\eta$,
as seen from (\ref{rhs}),(\ref{prs}). 
For small times however, $\rho_g$ and $p_g$ show strong deviations 
from this relation. In particular, $\rho_g$ as well as $p_g$ become 
{\it negative} for sufficiently small $\eta$. In the case $l=m=0$ 
(and therefore $\psi_2=0$) this is immediately seen from (\ref{rhs})
and (\ref{prs}). $\rho_g$ and $p_g$ are also negative for small
$\eta$ in more general spectra.

The main reason is that
the spectrum for wavelengths exceeding the local horizon scale at 
$\eta$ (corresponding to frequencies $k<1/\eta$) 
gives a negative contribution to the total values of energy density 
and pressure. If one goes back in time, the superhorizon modes 
ultimately dominate the spectrum (we always assume that the spectral 
density in the ultraviolet region drops sufficiently fast to ensure 
convergence for $k \rightarrow \infty$. Ultraviolet divergence is a 
problem of quantum theory and does not concern us here). 

Since the HF part of the spectrum satisfies a $\rho_ga^4 =
const$-law and is therefore not very interesting, we concentrate on 
the LF tail, which is essential for back reaction effects. 

\subsection{Back-reaction solutions in the regular low-frequency 
limit}

In the long-wavelength or LF limit $k \rightarrow 0$ we assume that 
the spectral density $f$ can be expanded in powers of 
$k$ around $k=0$. This assumption excludes a singularity at $k=0$, 
which in some cases can be interpreted as leading to a 
finite infrared contribution to $\rho$ and $p$ (Section V).
Writing down (\ref{al1}) explicitly requires to calculate the 
back reaction term $b$ and hence the integrals in (\ref{mom}).
In the sense of our approximation, this is trivial 
if the integration is assumed to extend to a maximal (but still 
small) $k=k_1$. Expanding also $\epsilon_0$, (\ref{al1}) leads to 
\begin{eqnarray}
 2qq'' - q'^{2} +4qq'\frac{a'}{a} && \nonumber  \\ 
+\frac{32\pi k_1^2\zeta}{3}q(-\frac{1}{5}k_1^2q^2 +\frac{1}{6}q'^2 
+qq'\frac{a'}{a})= 4\frac{\epsilon_{0}}{a^4}  \label{feq}
\end{eqnarray}
as lowest-order equation of a hierarchy of equations for the expansion
coefficients (we have written $q$ for $\lim_{k\to 0}f/a^2$ and again
$\epsilon_{0}$ for $ \lim_{k\to 0}\epsilon_0$).
Only this lowest-order approximation will be discussed. Since the 
solution of (\ref{al1}) can be written in the form (\ref{hom1}), 
it is appropriate to treat the case $\epsilon_{0}=0$.
Here $q=const$ is an obvious solution of (\ref{feq}),
provided we neglect the term proportional to $k_1^2$ in the 
bracket, which is small compared with the other terms in a LF
approximation. 
The effective energy density and pressure of the extreme LF
background follows as
\begin{equation}
\rho_g = \frac{3}{7}p_g = \frac{qk_1^5}{10 G a^2}. \label{eos}
\end{equation}
Note the $a^{-2}$ decay compared to the $a^{-4}$ decay of the energy 
density of the high-frequency radiation field (cf. \cite{Abramo2}). 
Thus the influence of the LF gravitational radiation field on the 
scale factor is small for sufficiently early times 
compared to other forms of radiation. The gravitational wave 
background considered here is alone not able to support a 
cosmological model: Since the time or scale factor dependence of 
energy density and pressure is already fixed, the two equations 
(\ref{g0}) and (\ref{gik}) are not compatible. But we may assume 
that a relativistic fluid with a $p_m= \rho_m/3$ equation of 
state is also present. This ensures compatibility at the price of
fixing the density ratio between fluid and waves and gives
\begin{eqnarray}
\rho_m = \frac{15a'^2-4a^2k_1^5q\pi}{40 G\pi a^4},\\
a'' + \frac{4}{5}\pi k_1^5qa =0. 
\end{eqnarray}
From the second relation the scale factor follows as
\begin{equation}
a = a_1 sin(l\eta), ~l^2=4\pi k_1^5q/5, \label{lfa}
\end{equation}
thus the energy density of the relativistic fluid changes as
\begin{equation}
\rho_m = \frac{k_1^5q(3-4\sin^2(l \eta))}
{10Ga_1^2 \sin^4(l \eta)}.
\end{equation}
It appears that the mere existence of a LF gravitational wave 
background could cause a reversal of the expansion, however 
small its energy density may be initially (that is, for times 
$\eta$ with $l\eta <<1$) compared to that of the relativistic
fluid, note that $\rho_g/\rho_m = \sin^2 (l\eta)/(3-4\sin^2(l\eta))$. 
For $l\eta > 0.8861$, $\rho_g$ exceeds $\rho_m$, for 
$\eta l >\pi/3$ 
the fluid energy density becomes formally negative through the 
interaction with the gravitational wave fields. 
The averaged Weyl tensor given by
\begin{equation}
\langle C_{0101}\rangle  = \langle C_{1212} \rangle =
\frac{16\pi a_1^2k_1^5q}{15}\sin^2(l\eta)
\end{equation}
vanishes at the singularity $\eta=0$ and is non-zero for $\eta >0$, 
suggesting that the presence of a LF component of gravitational 
radiation is not a coordinate effect. 

The case of a nonzero limit $\epsilon_0$ for 
$k \rightarrow 0$
is more complicated 
and seems not to allow an analytic solution.
As another example we treat the interaction of the LF background with
a $\Lambda-$term. It is common practice \cite{Lambd} to write the 
$\Lambda-$constant on the right-hand-side of the field equations, 
i.e. to consider a fluid with the components 
$\rho_m= \frac{\Lambda}{8\pi G},~ p_m=-\frac{\Lambda}{8\pi G}$.
Considering again the case  $\lim_{k\to 0}\epsilon_0= 0$, we arrive 
at (\ref{eos}) as before, but the 
field equations (\ref{g0}) and (\ref{gik}) lead now to an exponential 
increase or decrease of the scale factor (in the $\eta$-coordinate):
\begin{equation}
a = a_1 \exp{(\pm\eta\sqrt{4\pi k_1^5q/3}~)}.
\end{equation} 
As in the former case of a relativistic fluid, the "$\Lambda-$matter"
and the LF gravitational wave background are coupled since
\begin{equation}
q^3 = \frac{45 G\Lambda}{64 \pi^2k_1^{13}}
\end{equation}
is required for consistency.
The Weyl tensor components
\begin{equation}
\langle C_{0101}\rangle  = \langle C_{1212} \rangle 
= \frac{16\pi a^2k_1^5q}{15}
\end{equation}
are also finite (and nonzero) at $\eta =0$.

\section{Dominant Gravitational Wave Background}

Let us now assume a pure gravitational radiation universe. 
In the high frequency approximation all is already done: We do not 
have to face an appreciable back reaction term in the wave equation, 
the spectral shape of the waves is time-independent, the amplitude 
decreases only due to redshift effects, the equation of state can 
well be approximated by 
$\rho_g = 3 p_g$, and (\ref{g0}),(\ref{gik}) (with $\rho_m=p_m=0$)
give the Tolman radiation cosmos.

The presence of only LF components in the spectrum
does not allow a solution based on gravitational waves alone, 
thus the spectrum will not be restricted subsequently. 
We use the two equations (\ref{s1}) and
(\ref{s2}) for the frequency integrated quantities $g_4$ and $f_4$.
Adding (\ref{g0}),(\ref{gik}) with $\rho_m=0$ and $p_m=0$,
one has two further equations to determine the four unknown time
functions $f_2, f_4, g_4$ and $a$.  
We solve the last equations for $g_4$ and $f_4$ 
and use the result in (\ref{s2}) to find $f^{''}_2$. This
leads to    
\begin{eqnarray}
g_4 &=& \frac{1}{8\pi}(aa'' +3a'^2) +\frac{11}{3}\frac{a'^2}{a^2}f_2 
- \frac{4}{3}\frac{a'}{a}f_2',  \label{g}\\
f_4 &=& \frac{1}{8\pi}(-a a'' +3 a'^2) +\frac{10}{3}\frac{a'^2}{a^2} 
f_2 -\frac{5}{3}\frac{a'}{a}f_2',  \label{f4}\\
 f_2'' &=& \frac{1}{2\pi}a a'' 
+\bigl(2(1-2\zeta)\frac{a''}{a}
+\frac{2}{3}(1-6\zeta)\frac{a'^2}{a^2}\bigr)f_2 \nonumber  \\ 
&&+ \frac{2a'}{3a}f_2', \label{f2}\\ 
0 &=& 2\frac{a'}{a}\bigl((3\zeta +2)\frac{a''}{a}
+ (3\zeta -4)\frac{a'^2}{a^2}\bigr)f_2 \nonumber \\ 
&& + \bigl((\zeta -2)\frac{a''}{a} +(\zeta +4)\frac{a'^2}{a^2}\bigr)f_2' 
 \label{a}
\end{eqnarray}
This system may be studied in two cases, (i) assuming $\zeta=0$, that is,
no back reaction in the wave equation and (ii) full back reaction, 
$\zeta=1$. Furthermore one has to solve the generalized wave equation 
(\ref{we}) and check the compatibility of its solution with the moments
derived from (\ref{g})-(\ref{a}). 
Finally, if all this succeeds,
the energy density and pressure of the waves follow from the familiar 
equations
\begin{equation}
\rho_g = 3a'^2/(8\pi Ga^4), ~~p_g= (-2aa''+a'^2)/(8\pi Ga^4). \label{rp}
\end{equation}

\subsection{$\zeta=0$: Tolman universe}
With $\zeta =0$, (\ref{a})  
can be written as the product of two factors:
\begin{equation}
(2a'f_2 -af_2' )~(aa''-2 a'^2) = 0. \label{prod}
\end{equation}
Thus two different cases emerge,depending on which factor vanishes. 
If the first factor in (\ref{prod}) is zero, one obtains
with an integration constant $c$ (which may be gauged to zero)
\begin{equation}
f_2 = c a^2. \label{ff2}
\end{equation}
This relation is compatible with (\ref{f2}) if and only if $a''=0$, 
that is, if the scale factor has the time dependence $a=b\eta$ of 
the Tolman radiation cosmos \cite{Tolman}, with the energy density and 
pressure given by 
\begin{equation}
\rho_g = 3p_g = \frac{3}{8\pi Gb^2\eta^4}. \label{rt}
\end{equation}
From (\ref{g}, \ref{f4}) it follows that the other moments are
time-independent:
\begin{equation}
g_4= b^2(c+\frac{3}{8\pi}),~f_4 = \frac{3b^2}{8\pi}. \label{gf4}
\end{equation}
The average of the Weyl tensor component $C_{0101}$ or $C_{1212}$ 
comes out as the constant $2b^2$. The expressions for $f_2,f_4$ 
and $g_4$ found as solutions of differential equations must be 
compatible with those derived directly from the the spectral 
density (\ref{fall}). The three frequency dependent functions 
$l,m,n$ in (\ref{fall}) must therefore be chosen so that the 
moments $f_2,f_4,g_4$ have the time dependence which we have just 
derived. Self-consistency is only ensured if the functions $l,m,n$ 
exist. 
Working with real quantities, the spectral density 
$f$ is given by
\begin{equation}
f= 2(n+l\cos(2k\eta)-m \sin(2k\eta))
\end{equation}
in the radiation cosmos. As one verifies from the definition of
$g_4$ in (\ref{momg}) and of $f_0$ and $f_1$ in (\ref{mom}),
$g_4$ can be written
\begin{equation}
 g_4 = 4n_4 -f_4.  \label{gg}
\end{equation}

From the definitions of $f_2$ and $f_4$ in (\ref{mom}) one
obtains together with (\ref{ff2}) and (\ref{gf4})
\begin{eqnarray}
\int k^2(l \cos(2k\eta) - m \sin(2k\eta))dk &=&
 cb^2\eta^2/2 -n_2,\label{cond1} \\
\int k^4(l \cos(2k\eta) - m\sin(2k\eta))dk &=& -cb^2/4.\label{cond2}
\end{eqnarray}
(\ref{cond2}) can be obtained from
(\ref{cond1}) by differentiation with respect to $\eta$, so 
only (\ref{cond1}) is needed. All $k$-integrations considered 
so far run from $k=0$ to infinity. We formally extend $l(k)$ and 
$m(k)$ to negative values by
\begin{equation}
l(k) = l(-k) , ~m(k) = - m(-k) ~for~k<0. \nonumber
\end{equation}
This allows us to rewrite (\ref{cond1}) as complex Fourier transform
\begin{eqnarray} \label{fourier}&& 
{\cal F} [ \phi(k) | 2\eta ] \equiv \nonumber \\
&& \frac{1}{\sqrt{2\pi}}
\int^\infty_{-\infty} \phi(k)
\exp(2ik\eta)\,dk =  
\sqrt{\frac{2}{\pi}}(\frac{cb^2\eta^2}{2} -n_2)
\end{eqnarray}
applied to the complex valued function $\phi(k)=k^2(l(k)+im(k))$.
Since $\phi^*(k)=\phi(-k)$, the Fourier transform of $\phi(k)$
is real. To find $\phi(k)$, one has to invoke the Fourier inversion 
theorem. This requires to consider the right-hand-side of 
(\ref{fourier}) also for negative values of $\eta$. Since the right 
hand side is not absolutely integrable over the whole time
axis, $\phi(k)$ must be understood as generalized function (see, 
e.g., \cite{Sneddon}, \cite{Zemanian}, \cite{Folland} for a 
confirmation of the subsequent calculations). Extending the function 
space in this way, the Fourier inversion theorem remains valid. 
For instance, a polynomial in $\eta$ gives rise to the Dirac delta 
function $\delta(k)$ and its derivatives in Fourier space. One obtains
\begin{eqnarray}
k^2l(k)&=& -\frac{cb^2}{4}\frac{d^2 \delta}{d k^2}
-\frac{n_2}{4\pi}\delta(k)\\ k^2m(k)&=& 0.
\end{eqnarray}
We may also extend $n(k)$ to
negative values by $n(k)=n(-k)$ for $k<0$. The spectral components of
energy density $\rho(k,\eta)$ and pressure $p(k,\eta)$ are then 
symmetric functions of $k$, and the frequency integrated total 
density can be written 
$\rho= \frac{1}{2}\int_{-\infty}^\infty dk \rho(k,\eta)$, with a 
similar extension of the integration interval for the pressure. This 
allows to handle terms involving delta functions applying the usual 
rules for these functions. Replacing second derivatives of the
delta function using the formulae (cf.\cite{Sneddon}, the prime here
denotes the derivative with respect to $k$)
\begin{equation}
s(k) \delta''(k) = s''(0) \delta(k) -2s'(0)\delta'(k) +s(0)\delta''(k),
\end{equation}
one obtains for the spectral decomposition of the energy density and 
pressure from (\ref{rh1}), (\ref{pr1})

\begin{eqnarray}
a^4G\rho_g(k,\eta) &=& 2n(k)k^4 - \frac{7}{\eta^2}n(k)k^2
+\frac{7}{4\eta^2}cb^2\delta''(k) \nonumber \\&&
+(\frac{14}{\eta^2}n_2-cb^2)\delta(k),
\end{eqnarray}

\begin{eqnarray}
3a^4Gp_g(k,\eta) &=& 2n(k)k^4 - \frac{5}{\eta^2}n(k)k^2
+\frac{5}{4\eta^2}cb^2\delta''(k) \nonumber \\&&
+(\frac{10}{\eta^2}n_2-cb^2)\delta(k).
\end{eqnarray}

The spectral quantities can be integrated immediately 
to give the finite total values
\begin{equation}
a^4G\rho_g = 3a^4Gp_g = 2n_4 - \frac{b^2c}{2}
\end{equation}
in agreement with (\ref{gg}), showing the self-consistency 
of the calculation.

The singularities of the spectral decomposition $\rho_g(k,\eta)$ 
show that the infrared mode $k=0$ contributes a {\it finite} 
and time-dependent amount $Ga^4\rho_{ir} = 
\frac{7}{\eta^2}n_2 -\frac{cb^2}{2} $ to the 
total energy density. 
At superhorizon scales, more precisely at scales with 
 $k\eta < \sqrt{3.5}$, the spectral components become negative. 
Similar conclusions hold for the pressure, which becomes negative for 
$k\eta < \sqrt{2.5}$. 
The integrated values of energy density and pressure however stay 
always positive thanks to the infrared behaviour of their spectral 
components. 

\subsection{$\zeta=0$: de~Sitter scale factor }

The vanishing of the second factor in (\ref{prod}) gives the scale 
factor of the de~Sitter universe, $a=\frac{1}{H\eta}$. The density 
and pressure calculated from (\ref{rp}) are time-independent:
\begin{equation}
\rho_g = -p_g = \frac{3H^2}{8\pi G }. \label{rs}
\end{equation}
The differential equation for $f_2$, (\ref{f2}), has the general 
solution
\begin{equation}
f_2 = \frac{r_0}{\eta^2} +s_0 \eta^{7/3}
- \frac{3}{13\pi H^2} \frac{\ln|\eta|}{\eta^2}
\label{f2s} \end{equation}
with two constants $r_0$ and $s_0$, where $r_0$ may be gauged to 
zero. From (\ref{g},\ref{f4}) one obtains
\begin{eqnarray}
f_4 &=& \frac{65}{9}s_0 \eta^{1/3} 
- \frac{27}{104}\frac{1}{\pi H^2\eta^4}, \label{f4s} \\
g_4 &=& \frac{r_0}{\eta^4} + \frac{61}{9}s_0 \eta^{1/3}
- \frac{3}{13}\frac{\ln|\eta|}{\pi H^2 \eta^4}
+ \frac{33}{104}\frac{1}{\pi H^2\eta^4},  \label{gs}
\end{eqnarray}
and the Weyl tensor components are found as 
\begin{equation}
\langle C_{0101}\rangle =\langle C_{1212}\rangle=
\frac{728}{27}\pi s_0\eta^{1/3} - \frac{10}{13}\frac{1}{H^2\eta^4}.
\end{equation}
We have again to check the compatibility of the spectral density $f$
(defined by (\ref{fall})) with the time dependence of its moments 
$f_2,f_4,g_4$ as given by the last three equations.
The integral $f_2$ corresponding to the definition in (\ref{mom}) is
\begin{eqnarray}
f_2 &=& 2n_2 +\frac{2}{\eta^2}n_0 + 2\int^\infty_0k^2(l\cos(2k\eta)
-m\sin(2k\eta))\,dk \nonumber \\
& &+\frac{2}{\eta^2}\int^\infty_0(-l\cos(2k\eta)+m\sin(2k\eta))\,dk
\nonumber\\
& &-\frac{4}{\eta}\int^\infty_0k(m\cos(2k\eta)+l\sin(2k\eta))\,dk.
\end{eqnarray}
A similar expression holds for $f_4$. Straightforward calculation 
shows that the integrand of 
$g_4 =\int^\infty_0 k^4\hat{g}(k,\eta)\,dk$  defined by
(\ref{momg}), has the form ($x=k\eta$)
\begin{eqnarray}
\hat{g} &=& u \cos(2x) +v \sin(2x) + w, \label{guvw} \\
u &=& \frac{2}{x^4}(-l(x^4-3x^2+1)+2mx(x^2-1)), \nonumber \\
v &=& \frac{2}{x^4}( m(x^4-3x^2+1)+2lx(x^2-1)), \nonumber \\
w &=& 2n(1-1/x^2+1/x^4). \nonumber
\end{eqnarray}
Rewriting the just derived integral expressions  for $f_2$ and
$g_4$ as well as for $f_4$ as complex Fourier transforms, we 
obtain:
\begin{eqnarray}
f_2 &=& 2n_2+ \frac{2}{\eta ^2}n_0 + \psi_2-\frac{1}{\eta ^2}\psi_0
+ \frac{2i}{\eta}\psi_1, \label{ff2s} \\
f_4 &=& 2n_4+\frac{2}{\eta ^2}n_2 +\psi_4-\frac{1}{\eta ^2}\psi_2
+\frac{2i}{\eta}\psi_3, \label{ff4s}\\
g_4 &=& 2n_4-\frac{2}{\eta ^2}n_2 +\frac{2}{\eta ^4}n_0 -\psi_4
-\frac{2i}{\eta}\psi_3 \nonumber \\ &&
+\frac{3}{\eta ^2}\psi_2 +\frac{2i}{\eta ^3}\psi_1 
-\frac{1}{\eta ^4}\psi_0, \label{ggs}
\end{eqnarray}
where $\psi_j$ is a family of time functions defined by
\begin{equation}
\psi_j  = \int^\infty_{-\infty} k^j(l+im)\exp(2ik\eta)\,dk
\label{four}
\end{equation}
($\psi_2$ was already introduced in Section IV).
As in the Tolman case we have extended the functions $l,m,n$ to 
negative values of $k$ by $l(-k)=l(k)$, $m(-k)=-m(-k)$ and 
$n(-k)=n(k)$, thus the functions $\psi_j$ with even (odd) $j$ are 
real (pure imaginary). The members of a $\psi$-family are connected 
by differentiation and integration according to the rule
\begin{equation}
\psi_j' = 2i\psi_{j+1},\label{psi}
\end{equation}
which holds also if the $\psi_j$ are generalized functions.   
The aim is again to obtain the complex spectral density $l(k)+im(k)$ 
from any of the time functions $\psi_j$ by inverting the 
corresponding Fourier integral. We start solving (\ref{ff4s}) for 
$\psi_2$, where $f_4$ is replaced by the expression (\ref{f4s}). 
Using the rule (\ref{psi}) repeatedly, one obtains the differential 
equation 
\begin{eqnarray} \label{ps2}
\eta^2 \psi_2''-4\eta\psi_2'+4\psi_2 =&& \nonumber \\
8n_2 + 8n_4\eta^2
+\frac{27}{26 \pi H^2\eta^2} - \frac{260}{9}s_0\eta^{7/3} &&
\end{eqnarray}
for $\psi_2$. Its solution is given by (note that adding a solution 
of the homogeneous equation would give the wrong time dependence)
\begin{equation} \label{psi2s}
\psi_2 = 2n_2-4n_4\eta^2 + 13 s_0 \eta^{7/3} 
+ \frac{3}{52\pi H^2}\eta^{-2} \label{psitwo}
\end{equation}
If one member of a $\psi$-family is known, other can be found by 
differentiation and integration. Straightforward calculation shows, 
that the $\psi$-functions derived from (\ref{psi2s}) satisfy also 
(\ref{ff2s}) and (\ref{ggs}), if $f_2$ and $g_4$ on the 
left-hand-sides are substituted from (\ref{f2s}) and (\ref{gs}). 
We now apply the Fourier inversion theorem
to (\ref{four}) with $j=2$ and $\psi_2$ taken from (\ref{psitwo}).
This requires a continuation of the (real) function $\psi_2$ 
into the region $\eta < 0$. Here only the term proportional to $s_0$ 
requires more consideration, but this term contributes nothing to 
$\rho_g$ and $p_g$, we can therefore put the integration constant 
$s_0$ equal to zero.  One then obtains 
\begin{eqnarray}
k^2 l(k) & = & n_4\frac{d^2\delta(k)}{dk^2} +2n_2\delta(k) 
-\frac{3|k|}{26\pi H^2}, \\ k^2 m(k) & = & 0.
\end{eqnarray}
Again infrared modes enter the spectral density $l(k)$, but contrary 
to the Tolman case one is free to specify the spectral function 
$n(k)$. If $n(k)$ is chosen as zero, the delta function terms in $f$ 
and in the spectral decomposition of $\rho$ and $p$ vanish, but $f$ 
has still singular terms seen in an expansion around $k=0$:
\begin{equation}
f = \frac{3}{13 \pi H^2}(\frac{1}{k} + \frac{1}{\eta^2k^3}) 
    - \frac{2\eta^4}{39\pi H^2}k^3 + o(k^5) \label{expk}
\end{equation}
(we missed in section IVC the de~Sitter case, since we had 
excluded infrared singularities).
In spite of this singularity, density and pressure of the 
gravitational waves integrate to the finite constant values 
(\ref{rs}). Additional terms from a spectral function $n(k) \neq 0$ 
add nothing to the total values $\rho_g$ and $p_g$, thus the 
singular infrared ($k=0$) component in $\rho_g$ is cancelled 
by the integrated contribution of $n$-modes with $k\neq0$, the 
same holds for the pressure. 

It is easy to extend the calculation to account for a cosmological 
constant by adding matter terms $\rho_m = \Lambda/(8\pi G)$
and  $p_m = -\rho_m$ to the equations (\ref{g0},\ref{gik}).
Repeating the calculation at the beginning of this section,
one obtains (for $\zeta=0$) the same product  (\ref{prod}) as 
in the absence 
of a $\Lambda$ constant. The vanishing of the first factor gives a 
Tolman-de~Sitter model, where gravitational radiation has an 
equation of state (EOS) 
of the form $p_g=\rho_g/3$ and an energy density decaying 
as $a^{-4}$. The scale factor $a(\eta)$
can be expressed in terms of an elliptical integral.
 
More interesting is the model corresponding to the vanishing second 
factor, since here the de~Sitter scale factor follows. No new 
calculation is needed: In equation (\ref{rs}) we have to substitute 
for $\rho_g$ and $p_g$ the total values $\rho_g+\rho_\Lambda$ and 
$p_g+p_\Lambda$. In the calculations following this equation 
we must only replace the de~Sitter Hubble constant $H$ in all 
equations by $\tilde{H}=H/(1-64\pi^2G^2\Lambda/(3H^2))^{1/2}$. 
The de~Sitter expansion is generated by two independent "matter" 
sources, by a genuine $\Lambda$ constant as well as by a suitable 
spectrum of gravitational waves. We may have an arbitrary 
(but time-independent) mixture of both ingredients.
For sufficiently small $\Lambda$ gravitons dominate. 
If $\Lambda$ reaches the threshold $\Lambda^* = H^2/(64\pi^2 G^2)$, 
$\rho_g$ becomes zero and turns to negative values for still 
larger $\Lambda$, to ensure the same total energy density for 
a different composition.
 
We have not discussed in this article the origin of the
primordial wave spectrum, but it should be noted that 
the expressions for energy density and pressure introduced 
here may be of interest for concrete models. In the case of a 
quantum origin due to vacuum fluctuations in a de~Sitter cosmos 
\cite{quantum}, the produced gravitons can be described by a 
two-point correlation function, which for                  
Bunch-Davies vacuum \cite{Bunch} corresponds 
to a classical spectral density $f$ given by  
\begin{equation}
f_{BD} = \frac{\hbar G}{k\pi^2}(1+ \frac{1}{\eta^2 k^2}).
\end{equation}
$f_{BD}$ is obtained by comparing the quantum expectation values 
of bilinear terms in the metric (as given, e.g., in \cite{Allen2}) 
with the stochastic averages discussed here.
The spectral density $f_{BD}$ has the same infrared singularity  
as the expression (\ref{expk}) derived previously. Since the 
coincidence of both spectra holds only approximately for small $k$, 
the back reaction of the Bunch-Davies gravitons on the scale 
factor will change the scale factor\footnote{Similar 
conclusions follow from the important work by Tsamis and Woodard, 
who have discussed {\it quantum} gravity back reaction on an 
inflationary expansion rate in numerous papers \cite{Tsamis}.}. 
Again, in spite of the singularity of the spectral density $f_{BD}$, 
the integrated values of energy density and pressure require no 
infrared cut-off. If one introduces an ultraviolet cut-off at the 
frequency $k_1$ with $x_1=k_1\eta$, one obtains for energy density 
and pressure of the Bunch-Davies gravitons 
\begin{equation}
\rho_g = \frac{\hbar H^4}{4\pi^2}x_1^2(x_1^2-7),
~p_g = \frac{\hbar H^4}{12\pi^2}x_1^2(x_1^2+1).\label{ds}
\end{equation}
Considering limiting cases, for high frequencies $x_1\gg 1 $ 
follows the expected EOS $p = \rho/3$, for low frequencies 
$x_1 \ll 1$ one has $p = -\rho/3$ (together with $\rho <0$). 
The expressions (\ref{ds}) for $\rho_g$ and $p_g$ apply only for 
a de~Sitter scale factor. As a result of to back reaction, the local values 
of $\rho_g$ and $p_g$ change with the background geometry 
(thus they are, in a sense, not local): The lower the frequency, 
the stronger is the dependence of the effective EOS on the 
background gravitational field.

\subsection{$\zeta=1$: Full back reaction models}
Unfortunately, the approach described in the previous subsections
does not work in the real case $\zeta=1$. 
Even the first step, finding analytic solutions of (\ref{f2}),(\ref{a})
for $a,f_2$, has been so far unsuccessful. The back reaction function
can be written $b= \zeta (\frac{a''}{a} +\frac{a'^2}{a^2})$ 
for a dominant gravitational wave background, thus the differential 
equation  
for the wave amplitudes $h$ takes for $\zeta=1$ the unusual form  
\begin{equation}
h'' +h(k^2 +\frac{a''}{a} +2\frac{a'^2}{a^2}) =0.  
\end{equation}
This equation suggests that graviton creation persists in a 
wave dominated universe, provided the scale factor is different 
from $a \sim \eta^{1/3}$. Note that the latter scale factor is 
inconsistent with the system (\ref{f2}),(\ref{a}). For a treatment  
of the general case one has to resort to numerical calculations,
which will be discussed elsewhere.

\section{Final remarks}
\label{remarks}

We have seen that the stochastic back reaction equations 
treated here form, on the one hand, an apparently  self-consistent 
system of equations with interesting solutions.  On the other hand, 
it is not clear, how far the solutions deviate from true 
solutions of Einstein's field equations, either for some range 
of parameters or in some regions of space-time. We shortly 
discuss what could be done to clarify and to improve the situation.

One has to realize that {\it ensemble averages} are considered, 
which always differ from actual realizations of a random process. 
This is inherent to the method and cannot be changed, but one is 
able to say something more about statistical deviations from 
true solutions, if a Gaussian or some other process is assumed.

The main shortcoming of the approach is the use of a second-order 
approximation to general relativity, but improvements are 
possible. In principle, Monin and Yaglom's statistical treatment of 
nonlinear field theories works for arbitrary nonlinearities, if 
they are present in polynomial form. Only the technical complexity 
grows in a full treatment, since many higher-order 
correlation functions must be taken into account. Writing the 
Einstein field equations as
\begin{equation}
(R_{\mu\nu}-\frac{1}{2}g_{\mu\nu}R^{\rho\sigma}g_{\rho\sigma})g^3
=\kappa T_{\mu\nu}g^3,
\end{equation}
where $g$ is the determinant of the metric tensor, we see that
the left-hand sides consist of huge polynomials of maximal 
degree 12 in the (exclusively) covariant components of the metric 
tensor and its first and second derivatives. Taking the ensemble 
average of these expressions leads to correlation functions up to 
sixth order. If the random process $g_{\mu\nu}$ is Gaussian, 
the higher order correlation functions can be reduced to the 
second-order functions studied in this article. This would allow 
us to turn the back reaction equations into -in some sense - exact 
relations. Present discussions on a primordial 
stochastic gravitational wave background usually assume a 
{\it quantum} origin of gravitons, which are born out of zero-point 
vacuum fluctuations. The classical correlations discussed here 
are expected to be related to quantum mechanical expectation values, 
thus it seems natural to assume Gaussianity.
 

\end{document}